\documentclass{ws-procs9x6-cpt22}
\begin{document}

\newcommand{\refeq}[1]{(\ref{#1})}
\def\etal {{\it et al.}}
\newcommand{\mbf}{\mathbf}
\newcommand{\gv}[1]{\ensuremath{\mbox{\boldmath$ #1 $}}}

\title{LIV in matter  }

\author{L.F.\ Urrutia }

\address{High Energy Physics Department, Instituto de Ciencias Nucleares, Universidad Nacional Aut\'onoma de M\'exico,\\
Ciudad de M\'exico, 04510, M\'exico}

\begin{abstract}
We summarize  recent work dealing with the characterization of effective actions giving the electromagnetic response of 
some topological materials, arising from their microscopic structure. The case of weakly tilted Weyl semimetals in the limit of zero temperature, but nonzero chemical potential, is presented  as a subset of a specific choice of terms in the fermionic sector of the SME.  
\end{abstract}
\bodymatter
\section{Introduction}
The extension of the constant  background tensors in the SME to spacetime dependent variables \cite{KLP} yields new possibilities of studying Lorentz invariance violations (LIV) which might result in  effective theories describing  macroscopic behavior of matter. Here we focus on the CPT-odd sector of the photon sector  of the SME where we promote the standard constant vector $k_{AF}^\mu$ to $\partial^\mu \theta(x)$. The resulting coupling has been  previously introduced in axion electrodynamics, where $\theta(x)$ is the axion field. \cite{SIKIVIE}. The nondynamical version of this theory  has  gained recent revival  in the description of the electromagnetic (EM) response of topological matter such as  topological insulators (TIs) and  Weyl semimetals (WSMs), for example. In these  cases, the LIV parameters are provided by the microscopic model of the matter in question and a crucial difference with the study  of LIV in the fundamental  interactions arises: in the former approach there is an underlying microscopic theory providing a Lorentz symmetry breaking pattern, while this
possibility is  missing in the second case  because  an established unified theory admitting LIV is still lacking. This is particularly notorious in some  calculation of radiative corrections  in the particle case yielding finite but indeterminate results, \cite{ALFARO} which nevertheless can be fixed in the matter calculation precisely due to the underlying theory.   This basic theory  further motivates the challenge of  
 determining how an effective macroscopic response arises from the microscopic  interactions describing the material. The analogy with the studies of LIV in the fundamental interactions is enhanced because the linearized approximation of the lattice Hamiltonians of TI's and WSMs close to the Fermi energy can be formulated in terms of massless chiral fermionic quasiparticles. This similarity brings about  the presence of anomalies  together with the question of their relevance in the construction of the corresponding effective actions.\cite{AGALU}
\section{Nondynamical axion electrodynamics ($\theta$-ED)}\label{sec1}
Perhaps the simplest way of introducing this modification of standard electrodynamics is by recalling  Maxwell equations in a linear nondispersive media in terms of the vectors $\mbf{D, H, E, B}$  in standard notation. To be complete they require constitutive relations $\mbf{D}=\mbf{D(E,B})$  and $\mbf{H}=\mbf{D(E,B})$ which define the medium to be considered. In this case we take 
$
\mathbf {D} = \varepsilon \mathbf {E} - \frac {\theta \alpha} {\pi} \mathbf {B}$ and $
\mathbf {H} = \frac {1} {\mu} \mathbf {B} + \frac {\theta \alpha} { \pi}
\mathbf {E}$, 
where $ \alpha \simeq 1/137 $ is the fine 
structure constant. The new ingredient here is the magnetoelectric polarizability $\theta(x)$ (MEP)
which is an additional parameter of the medium in the same footing as $\epsilon(x)$ and $\mu(x)$, i.e., the axion field is nondynamical. The resulting equations are the standard Maxwell equation in a medium having the additional field dependent  sources (in Gaussian units and $c=1$)  
\begin{eqnarray}
&&\rho_\theta= \frac{\alpha }{4 \pi^2 }\nabla
\theta \cdot \mathbf{B}, \qquad \mathbf{J}_\theta=-\frac{\alpha}{4 \pi^2}\Big(\nabla \theta \times \mathbf{E} + \frac{\partial \theta }{\partial t}\mathbf{B}\Big).
\label{aba:eq2}
\end{eqnarray} 
The usual  Maxwell action plus the additional  coupling $-\frac{\alpha}{4 \pi^2} \theta(x) \mbf{E}\cdot \mbf{B}$ reproduces the above equations. We remark that $\mbf{E}\cdot \mbf{B}$ is proportional to the Pontyagin density $\epsilon^{\alpha\beta\mu\nu} F_{\alpha\beta} F_{\mu\nu}$, which precludes  topological properties of the system. A most remarkable consequence of Eqs. \refeq{aba:eq2} is the so called magnetoelectric effect (MEE),  encapsulated in   the 
additional sources 
$\rho _{\theta }$ and $\mathbf{J}_{\theta }$. They summarize the ability of the magnetic (electric) fields to produce charge (current) densities, respectively. 
 $\theta$-ED describes the EM response of materials such as:  magnetoelectrics ($\theta$ piecewise constant but arbitrary), TIs ($\theta$ piecewise constant but quantized) and chiral matter ($\theta=b_\mu x^\mu$, with $b_\mu$ constant), among others.
\section{Radiation in $\theta$-electrodynamics}
The static limit of $\theta$-ED in the piecewise constant realization of the MEP has been thoroughly studied in numerous references summarized in Ref. \refcite{CAP}, either by the method of images or by constructing the Green's function (GF) of the system.  As a consequence of the MEE the GF turns out to be a matrix operator.  This method can be extended to the time dependent case and it is particularly useful in the far field approximation describing radiation. The evaluation of the GF in the radiation zone required to construct the electric an magnetic fields defining the Poynting vector involves integrals of highly oscillating functions, which can be  dealt   either with  the stationary phase approximation or with the steepest descent method in order to  obtain analytical results.  A remarkable result is the reversed Cherenkov radiation (CHR) in naturally existing materials found when a charge at constant velocity $v> c/n$ impinges  perpendicularly to  the planar interface between two TIs with the same permittivity $\epsilon=n^2$. \cite{FTU} In this case the angular distribution of the radiation includes a term $\delta(1-vn|\cos \theta|)$ which allows the contribution of angles $\pi/2<\theta <\pi $, thus yielding an additional  radiation cone in the backward direction with respect  to the incident charge. The power output in the reversed direction is highly suppressed with respect to the forward direction  by a factor of the order of $ (\theta\alpha/n)^2$, with  $\theta$ being the difference between  the MEPs of the two media. Reversed CHR has been observed only in metamaterials with negative permeability and permittivity. \cite{METAMAT} A similar approach is used in when dealing with radiation in chiral matter and the case of CHR also yields interesting results. \cite{BARREDO} 
\section{The effective action for  weakly tilted Weyl semimetals}
An important issue in  the study of LIV in matter is the construction of the macroscopic EM response  starting from the corresponding microscopic Hamiltonian coupled to the EM field. We illustrate this point taking as a model  the Hamiltonian 
\begin{eqnarray}
H _{\chi} (\gv{p}) = {\gv v}_\chi \cdot ({\gv p}   +  \chi {\gv {\tilde b}})  -   \chi {\tilde b}_{0}  + \chi v _{F} {\gv \sigma} \cdot ({\gv p}  + \chi {\gv {\tilde b}}) , \label{aba:eq3}
\end{eqnarray}
which describes a Weyl semimetal with two  3D band crossings (Weyl points) of chirality $\chi = \pm 1$
separated by   {$2 {\gv {\tilde b}}$}  and {${2\tilde b} _{0}$} in momentum and energy, respectively.
Equation \refeq{aba:eq3} corresponds to  the linearized approximation of the  lattice Hamiltonian close to the Fermi energy and clearly shows the appearance of Weyl excitations around each Weyl point. 
Here $v_{F}$ is the isotropic Fermi velocity at each band crossing, ${\gv \sigma} = (\sigma _{x} , \sigma _{y} , \sigma _{z})$ is the triplet of spin-$1/2$ Pauli matrices,  ${\gv v} _{\chi}$ is the tilting parameter and ${\gv p}$ is the momentum. In the Weyl (chiral) representation of the matrices $\gamma^\mu$,  the Hamiltonian   \refeq{aba:eq3} can be shown to emerge from the Lagrangian density ${\cal L}= \bar{\Psi} \left( \Gamma ^{\mu} i \partial _{\mu} - M \right) \Psi$ with 
\begin{equation}
\Gamma ^{\mu }= c^{\mu }{}_{\nu }\gamma ^{\nu }+d^{\mu
}{}_{\nu }\gamma ^{5} \gamma ^{\nu },\qquad M=a_{\mu }\gamma ^{\mu }+b_{\mu
}\gamma ^{5} \gamma ^{\mu },
\label{aba:eq4}
\end{equation}
which we recognize as a partial contribution from the fermionic sector in the minimal QED extension of the SME. \cite{datatables} We follow the conventions of Ref. \refcite{BD}. The term $c^\mu{}_\nu$ already includes the  $\delta^\mu_\nu$ contribution corresponding to the free Dirac action and we impose the restriction $\Gamma^0=\gamma^0$. Let us remark that the choices in Equation \refeq{aba:eq4} include also anisotropy in the Fermi velocity at each node (not shown in Equation \refeq{aba:eq3}). We emphasize that the LIV parameters in \refeq{aba:eq4} can be determined as  functions of the parameters in the microscopic Hamiltonian
in each case. 
Coupling the fermionic action arising from Equation \refeq{aba:eq4} to the EM field and using standard field theory methods we obtain the  effective EM response codified in the Lagrangian density ${\cal L}(p)=\frac{1}{2} \, A _{\mu} (-p) \, \Pi ^{\mu \nu }(p) \, A_{\nu} (p)$ in momentum space. We  require the calculation of the CPT-odd contribution to the vacuum  polarization tensor (VPT) $\Pi_{\rm A}^{\mu\nu}$, which has a well-known expression  \cite{BD}
in terms of the exact fermion propagator  $S(k) = i / ( \Gamma ^{\mu} k _{\mu} - M )$  including LIV modifications.  
The final result is that the EM effective action  will keep the form of $\theta$-ED for chiral matter, with all the parameters entering through a new vector ${\cal B}_\lambda$ to be determined.
The link between the relevant expressions is 
\begin{equation}
\Pi_{\rm A}^{\mu \nu} (p) = -i ({e ^{2}}/{2 \pi ^{2}}) \, {\cal B} _{\lambda} p _{\kappa} \epsilon ^{\mu \nu \lambda \kappa} , \quad \theta(x) \,\, \rightarrow \,\, \Theta (x) = 2 {\cal B} _{\lambda}  x ^{\lambda}. \label{aba:eq7} 
\end{equation}
The  action resulting  from the choice \refeq{aba:eq4} is chiral invariant, which simplify enormously the calculations allowing the splitting of the VPT into the chiral contributions $\Pi_{{\rm A}, \chi}^{\mu\nu}, \, \chi=\pm 1$, 
\begin{equation}
\Pi _{A, \, \chi }^{\mu \nu }(p) = - 2 {\chi} e ^{2} (\det m_{\chi})(m_\chi^{-1}) ^{\rho} {}_{\lambda} \epsilon ^{\mu \nu \lambda \kappa} p _{\kappa} \, I _{\rho} (C^{(\chi)}).  
\end{equation}
Here we have  $(m_\chi)^\mu{}_\nu= c^\mu{}_\nu - \chi\, d^\mu{}_\nu$ together with  $C_\rho^{(\chi)}
=a_\rho -\chi\, b_\rho$, and 
\begin{eqnarray}
 I _{\rho} (C^{(\chi)}) = \int  \frac{d ^{4} k }{\left( 2 \pi \right) ^{4}} \, g^{(\chi)} _{\rho} (k _{0} , {\gv k}) \, , \quad  g^{(\chi)} _{\rho} (k _{0} , {\gv k}) = \frac{ \left( k^{(\chi)} - C^{(\chi)} \right) _{\rho} }{ \left[ \left( k^{(\chi)} - C^{(\chi)} \right) ^{2}\right] ^{2} }, 
\end{eqnarray}
with ${k^{(\chi})}_\mu={k}_\alpha (m_\chi)^{\alpha}{}_\mu$. In order to extend the phenomenological applications of our approach we consider the case of nonzero chemical potential $\mu$ in the limit where the temperature $T$ goes to  zero. To this end we adopt the imaginary time  regularization where   $\int dk_0 \,  f(k_0)$ is replaced by the Matsubara sum  $\sum_n 2\pi i T \, f(k_0) $ with the replacement $k_0 \,\rightarrow \, i(2n+1) \pi T + \Lambda$. Here $\Lambda$ is the chemical potential measured from the band-crossing points, i.e,
$\Lambda=\mu-E_{\chi}({\gv{p}}_{\chi})$, where ${\gv{p}}_{\chi}$ is the
location in momentum of the node with chirality $\chi$, and
$E_{\chi}({\gv{p}}_{\chi})$ the corresponding energy. The required limit $T \rightarrow 0$ splits into the  contributions ${\cal B}^{(1)}_\lambda$ and $ {\cal B}^{(2)}_\lambda$ to ${\cal B}_\lambda$ in Eq.\ \refeq{aba:eq7}
\begin{eqnarray}
&&{\cal B}_\lambda^{(1)}=-\frac{1}{2}\sum_{\chi} \chi C_\rho^{(\chi)}\, (m_\chi^{-1})^\rho{}_\lambda, \qquad 
{\cal B}_\lambda={\cal B}^{(1)}_\lambda + {\cal B}^{(2)}_\lambda, \nonumber \\
&&  {\cal B}_\lambda^{(2)}=- \sum_\chi \chi \Lambda_\chi \, N_\chi \, {\cal V}_i^{(\chi)}\, (m_\chi^{-1})^i{}_\lambda, \quad {\cal V}^{(\chi) i}=(m_\chi^{-1})^i{}_j (m_\chi)^j{}_0 \label{aba:eq9} \\
&& \Lambda_\chi=\mu-\Big( {\cal V}_i^{(\chi)} C_i^{(\chi)} +C_0^{(\chi)} \Big),\,\,  N_\chi= \frac{1 }{2|\gv{\mathcal{V}}^{(\chi)}|^3}\Big(|\gv{\mathcal{V}}^{(\chi)}|-\mathrm{arctanh}
    (|\gv{\mathcal{V}}^{(\chi)}|)\Big), \nonumber 
\end{eqnarray}
where $|\gv{\mathcal{V}}^{(\chi)}|$ is the modulus of the vector
${\cal V}^{(\chi)i}$. Equations \refeq{aba:eq9} provide the full vector ${\cal B}_\lambda$ which defines the EM effective action corresponding to the fermionic sector of the SME  stemming from the choice in Eq. \refeq{aba:eq4}.\cite{AGLU} Applications of this EM response will be considered in a following work.
For further implementation of the methods and techniques motivated  by the SME into the realm of topological matter see  Refs.\ \refcite{MCG}.  
\section*{Acknowledgments}
L.F.U. acknowledges support from the project 
CONACyT-CF-2019-428214.

\end{document}